\input harvmac

\def\Title#1#2{\rightline{#1}\ifx\answ\bigans\nopagenumbers\pageno0
\else\pageno1\vskip.5in\fi \centerline{\titlefont #2}\vskip .3in}

\font\caps=cmcsc10

\noblackbox
\parskip=1.5mm

  
\def\npb#1#2#3{{\it Nucl. Phys.} {\bf B#1} (#2) #3 }
\def\plb#1#2#3{{\it Phys. Lett.} {\bf B#1} (#2) #3 }
\def\prd#1#2#3{{\it Phys. Rev. } {\bf D#1} (#2) #3 }

\def\pr#1#2#3{{\it Phys. Rep. } {\bf #1} (#2) #3 }

\def\bb#1{{\tt hep-th/#1}}


\def\CL{{\cal L}}   
   
\def\CM{{\cal M}} 
\def\CN{{\cal N}}


\def\dj{\hbox{d\kern-0.347em \vrule width 0.3em height 1.252ex depth
-1.21ex \kern 0.051em}}

\def\half{{1\over 2}\,}

\def\Tr{{\rm Tr\,}}

\def\ket{\rangle}
\def\bra{\langle}

\def\tphi{\widetilde \phi}
\def\tH{\widetilde H}
\def\tpsi{\widetilde \psi}
\def\tZ{\widetilde Z}
\def\tz{\tilde z}

\lref\regk{S. Elitzur, A. Giveon and D. Kutasov, 
 \plb{400}{1997}{269,} \bb{9702014.}}
\lref\regkrs{S. Elitzur,
 A. Giveon, D. Kutasov, E. Rabinovici and A. Schwimmer, 
{\it ``Brane Dynamics and N=1 Supersymmetric Gauge Theory",}
\bb{9704104.}}   
\lref\rvaf{M. Bershadsky, A. Johansen, T. Pantev, V. Sadov and
C. Vafa, {\it ``F-theory, Geometric Engineering and $N=1$ dualities",}
\bb{9612052\semi} 
C. Vafa and B. Zwiebach, {\it ``$N=1$ Dualities of
$SO$  and $Usp$ Gauge Theories and T-Duality of String Theory",}
\bb{9701015\semi}
C. Vafa and H. Ooguri, {\it ``Geometry of $N=1$ dualities
in Four Dimensions",} 
\bb{9702180.}}
\lref\rhw{A. Hanany and E. Witten, \npb{492}{1997}{152,} \bb{9611230.}} 
\lref\rdeff{P.C. Argyres, M.R. Plesser and N. Seiberg, \npb{471}{1996}
{159,}  \bb{9603042.}} 
\lref\rpol{J. Polchinski, {\it ``TASI lectures on D-branes'',} 
\bb{9611050.}}
\lref\rdm{M.R. Douglas and G. Moore, {\it ``D-branes, Quivers and 
ALE Instantons'',} \bb{9603167}.}
\lref\rwinst{E. Witten, \npb{460}{1996}{335,} \bb{9510135\semi} 
\npb{460}{1996}{541,} \bb{9511030.}}
\lref\rdoug{M.R. Douglas, {\it ``Branes within Branes'',} \bb{9512077\semi} 
{\it ``Gauge Fields and D-branes'',} \bb{9604198.}}
\lref\rvafi{C. Vafa, \npb{463}{1996}{435,} \bb{9512078.}}
\lref\rahh{O. Aharony, A. Hanany, {\it ``Branes, Superpotentials
and Superconformal Fixed Points",} 
\bb{9704170.}}
\lref\rbdh{J.H. Brodie and A. Hanany, {\it `` Type-IIA Superstrings,
Chiral Symmetry and $N=1$ 4D Gauge Theory Duality",} 
\bb{9704043.}} 
\lref\rahz{A. Hanany and A. Zaffaroni, {\it ``Chiral Symmetry from
Type-IIA Branes",} 
\bb{9706047.}}
\lref\rclf{N. Evans, C.V. Johnson and A.D. Shapere, {\it ``Orientifolds,
Branes, and Duality of 4D Gauge Theories",} \bb{9703210.}} 
\lref\rbarb{J.L.F. Barb\'on, \plb{402}{1997}{59,} \bb{9703051.}} 
\lref\rberk{J. de Boer, K. Hori, H. Ooguri, Y. Oz and Z. Yin, \npb{493} 
{1997}{148,} \bb{9702154.}} 
\lref\rtheis{A. Brandhuber, J. Sonnenschein, S. Theisen and S. 
Yankielowicz, {\it ``Brane Configurations and 4D Field Theory Dualities",} 
\bb{9704044.}} 
\lref\rwittenM{E. Witten, {\it ``Solutions Of Four-Dimensional Field
Theories via M Theory",} \bb{9703166.}} 
\lref\rmberk{K. Hori, H. Ooguri and Y. Oz, {\it ``Strong Coupling
Dynamics of Four-Dimensional $N=1$ Gauge Theories from M Theory
Fivebrane",} \bb{9706082.}}
\lref\rkapl{A. Brandhuber, N. Itzhaki, V. Kaplunovsky, J. Sonnenschein
and S. Yankielowicz, {\it ``Comments on the M Theory Approach to
$N=1$ $SQCD$ and Brane Dynamics",} \bb{9706127.}} 
\lref\rrut{O. Aharony, A. Hanany, K. Intriligator, N. Seiberg and
M.J. Strassler, {\it ``Aspects of $\CN =2$ Supersymmetric Gauge
Theories in Three Dimensions",} \bb{9703110.}} 
\lref\rwitqcd{E. Witten, {\it ``Branes And The Dynamics Of QCD",} 
\bb{9706109.}} 
\lref\rmqcd{A. Hanany, M.J. Strassler and A. Zaffaroni, {\it ``Confinement
and Strings in MQCD",} \bb{9707244.}} 
\lref\raffl{I. Affleck, \npb{191}{1981}{429.}} 
\lref\rtof{G. 't Hooft, \prd{14}{1976}{3432.}}
\lref\rfy{Y. Frishman and S. Yankielowicz, \prd{19}{1979}{540.}} 
\lref\ritep{V.A. Novikov, M.A. Shifman, A.I. Vainshtein, V.B. Voloshin
and V.I. Zakharov, \npb{229}{1983}{394\semi} 
V.A. Novikov, M.A. Shifman, A.I. Vainshtein and V.I. Zakharov, \npb{260}
{1985}{157.}}
\lref\rrev{D. Amati, K. Konishi, Y. Meurice, G. Rossi and G. Veneziano,
\pr {162}{1988}{169.}}                            
\lref\rdorey{N. Dorey, V.V. Khoze and M.P. Mattis, \prd {54}{1996}{2921,} 
\bb{9603136\semi} 
\prd{54}{1996}{7832,}
\bb{9607202.}}
\lref\rgomez{C. G\'omez and R. Hern\'andez, {\it ``M and F theory instantons,
N=1 supersymmetry and fractional topological charges'',} \bb{9701150.}}
\lref\ryi{K. Lee and P. Yi, {\it ``Monopoles and instantons on partially
compactified D-branes'',} \bb{9702107.}}
\lref\rangles{M. Berkooz, M.R. Douglas and R.G. Leigh, \npb{480}{1996}{265,}
\bb{9606135.}}  
\lref\rads{I. Affleck, M. Dine and N. Seiberg, \npb{241}{1984}{493.}}
\lref\rlands{K. Landsteiner, E. Lopez and D.A. Lowe, {\it ``$N=2$ 
Supersymmetric Gauge Theories, Branes and Orientifolds",} \bb{9705199.}}
\lref\rbrand{A. Brandhuber, J. Sonnenschein, S. Theisen and 
S. Yankielowicz, {\it ``M Theory and Seiberg-Witten Curves: Orthogonal
and Symplectic Groups",} \bb{9705232.}} 
\lref\rwihiggs{E. Witten, {\it ``On The Conformal Field Theory Of The Higgs 
Branch'',} \bb{9707093.}}


\line{\hfill CERN-TH/97-196}
\line{\hfill IFUM-576/FT}
\line{\hfill {\tt hep-th/9708041}}
\vskip 1.2cm

\Title{\vbox{\baselineskip 12pt\hbox{}
 }}
{\vbox {\centerline{D0-Branes, Constrained Instantons and}
\vskip10pt 
\centerline{D=4 Super Yang-Mills Theories }
}}

\vskip0.6cm

\centerline{$\quad$ {\caps J. L. F. Barb\'on~$^1$ and A. Pasquinucci~$^{1,2}$
 }}
\vskip0.8cm

\centerline{{\sl $^1$ Theory Division, CERN}}
\centerline{{\sl 1211 Geneva 23, Switzerland}}
\centerline{{\tt barbon@mail.cern.ch, pasquinu@mail.cern.ch}}

\vskip0.3cm

\centerline{{\sl $^2$ Dipartimento di Fisica, Universit\`a di Milano }}
\centerline{{\sl and INFN, sezione di Milano}}
\centerline{{\sl via Celoria 16, 20133 Milano, Italy}}

\vskip 1.0in

\noindent{\bf Abstract:} We consider in more detail the role of D0-branes
as instantons in the construction  of $SU(N)$ Super 
Yang-Mills and Super QCD
theories in four space-time dimensions 
with D4, D6 and NS-branes. In particular, we show how the D0-branes
describe both the exact and constrained instantons and reproduce the 
correct pattern of lifting of zero modes on the various branches of 
these models. 




\Date{August 1997}



\newsec{Introduction}
In this paper we consider the by now standard construction
of four-dimensional super Yang-Mills (SYM) theories with 
Dirichelet branes (D-branes) \refs{\rhw,\regk} (see also 
\refs{\rberk,\rbarb,\rclf,\rbdh,\rtheis,\regkrs,\rahh,\rahz})    
and discuss how the world-line dynamics 
of D0-branes bound to D4-branes describes field theory SYM 
instantons (a more geometrical approach was developed in refs.\ 
\refs\rvaf).   

The basic construction of SYM instantons on a D$p$-brane world-volume, 
is as bound states with D$(p-4)$-branes \refs{\rwinst,\rdoug,\rvafi,\rdm}. 
In the case of the construction of ref.\ \regk, the $SU(n_c)$ SYM theory
is described by $n_c$ D4-branes, so we must consider D0-branes
bound to D4-branes. The  five-dimensional
world-volume of the D4-branes is then compactified 
to the four-dimensional physical space and, to make a bound D0-brane 
describe an instanton in the four dimensional physical space, the 
D0-brane euclidean-time world-line must lie along the compactified 
direction of the D4-brane.  Although the
details of such reduction  
depend on the amount of supersymmetry  ($N=4,2,1,0$) that we want
to preserve in space-time, the basic properties of the
instanton moduli space only depend on the geometry of D0-D4 branes
bound states, which we review in this introduction.      

The collective dynamics of the instantons is specified by a point
action and an appropriate measure over instanton moduli space. In
the present D0-branes construction, both ingredients appear as the
dimensional reduction down to zero dimensions of the D0-brane 
one-dimensional collective dynamics. In particular, the instanton 
moduli space as a hyper-Kahler quotient appears as a  
Higgs branch of the world-line theory of the D0-branes.
The basic terms in the world-line action are 
the static Born-Infeld action of the D0-brane
\eqn\bi{
S_{BI} = {\tau_0 } \int \sqrt{{\rm det}\, h_{\rm induced}}\ \longrightarrow\
M_{D0} \int d\tau,}
and a source term coming from the Chern-Simons couplings on
the D4-branes world-volume 
\eqn\source{S_{\rm source} = i {\mu_4 \over 2}
 \int_{\Sigma_{4+1}} A_{D0}^{RR} \,(2\pi\alpha')^2 \,
F\wedge F,}  
which relates the RR photon of the type-IIA string theory
to the theta parameter of the four-dimensional
instanton.\foot{For a constant one-form $A_{D0}^{RR}$, the five
dimensional integral \source\ is localized 
 on the D0-brane world-line. If the proper time integral
extends in the interval $\tau \in [0,L]$, we have $S_{\rm source} =
-i {\theta \over L} \int d\tau $.} There are also        
non-trivial gauge interactions specifying the dynamics of 0-0 strings
and 0-4 strings in the general case where we have $k$ D0-branes 
bound to $n_c$ D4-branes.  
The 0-0 sector is just the dimensional reduction down to one
dimension of ten-dimensional SYM with $U(k)$ gauge group, corresponding to
a system of $k$ D0-branes, and has $N=4$ supersymmetry in four-dimensional
 notation. The
0-4 sector breaks half of these unbroken 16 supercharges, leading
to an action for the 0-4 strings with $N=2$ supersymmetric couplings.
To be more explicit, let us parametrize the relevant degrees of freedom
in terms of a T-dual configuration of D3-branes inside D7-branes of
type IIB string theory. We will take the D3-branes world-volume to lie 
in the $(x^6, x^7, x^8, x^9)$ plane, with $x^6$ euclidean time, and 
the D7-branes to extend further in the $(x^0, x^1, x^2, x^3) $, 
all euclidean, directions. 
(These conventions for the axes follow from the conventions of 
ref.\ \regk\ where the world-volume of the D4-branes is along the 
$(x^0, x^1, x^2, x^3, x^6) $ directions, and the request that the 
euclidean-time world-line of the D0-branes is in the compactified Neumann 
direction of the D4-branes, that is along $x^6$.) 
Using $N=1$ superspace notation, the 0-0 sector contains a vector
superfield built from the gauge field $(A_6, A_7, A_8, A_9)$ on the
D3-brane world-volume, and the ``gluino" superpartners, and also three
chiral superfields in the adjoint of the $U(k)$ gauge symmetry: $W,
Z, \tZ$, whose bosonic components are the complex combinations
$X^4 +iX^5, X^0 + iX^1 , X^2 +iX^3$ respectively, and correspond to
the transverse motion of the D3-brane. Upon dimensional reduction to the
D0-brane world-line, we will gauge-fix $A_6$ to zero, and consider only 
configurations with $A_7 =0$, so that the remaining components of
the gauge field can be assembled into another chiral superfield
$A_8 +iA_9 = X_8 + iX_9 = V$. The fermionic components of these superfields
exhaust the $16k$ fermionic components of the ten dimensional $U(k)$
SYM theory.
If we denote the fields $\Phi_{00}=(W, Z, {\widetilde Z})$ collectively by
$\Phi_i, \,i=1,2,3$, the 0-0 action is defined by the following
$N=4$ superpotential:
\eqn\oosup
{{\cal W}_{0-0} = {\sqrt{2} \over  g_s}  \, {1\over 3!}\,
\epsilon_{ijk}
\,\Tr\, \Phi_i [\Phi_j, \Phi_k] \ .}   

The 0-4 sector contributes hypermultiplets in the fundamental and
anti-fundamental representations of the $U(k)$ symmetry, with the
D4-brane gauge indices $a,b,\ldots=1,\ldots,n_c$ playing here the role 
of ``flavour". 
The 0-4 strings contribute a ``D04-quark superfield" $\Phi_{04} =
 H = (\phi_h , \psi_h)$,
and the orientation-reversed 4-0 sector gives the D40-anti-quark
$\Phi_{40} = \tH = (\tphi_h , \tpsi_h)$. 
By coupling these hypermultiplets to the previous 0-0 chiral superfields
in the adjoint of $U(k)$, we break $N=4$ supersymmetry to $N=2$ and 
the $SO(6)$ global R-symmetry down to $SU(2)_R$. 
Indeed the D04-quark and D40-anti-quark 
fields $(H, \tH)$ couple 
in the superpotential to only one of the three 0-0 superfields. 
The 0-0 superfield which is singled out in this way is the 
$W$ superfield according to its four-dimensional
 field theory interpretation. The 
corresponding superpotential reads
\eqn\ocsup
{{\cal W}_{0-4} = {\sqrt{2}\over g_s}  \, ( \tH W H -
H W' \tH ), }
where we have introduced a mass term inherited from the superpotential
in the D4-branes, i.e.\ the 4-4 strings, coupling the $SU(n_c)$ adjoint
$W'_{44} = X'_4 + iX'_5$ giving the position of D4-branes. The $N=2$ 
supersymmetry
on the D0-brane world-line implies that $[W', W'^{\dagger}]=0$, i.e.\ the
$W'$ field lies along the flat directions of the D4-brane theory, and can
be taken as diagonal matrices by a $SU(n_c)$ rotation\foot{The brane
construction of the full gauge theory produces naively a $U(n_c)$ group.
Only the $SU(n_c)$ subgroup is relevant to the instanton discussion,
of course. In fact, on the Coulomb branch of the D4-branes world-volume,
the extra $U(1)$ is frozen, since fluctuations of the trace of $W'$
have infinite norm \refs\rwittenM.}   
$W'^b_a = w'_a \,\delta^b_a$.    

The instanton moduli space is then obtained as a maximal Higgs phase  
of this theory on the D0-brane world-line, where we consider the most
general vacuum expectation values of the hypermultiplets $H, \tH$ and
the adjoint fields $Z, \tZ$. In this paper we will be mainly concerned
with properties of the single instanton sector, or equivalently, of the
dilute gas limit of instantons. 
In this case the  overlap  between the instantons
 can be neglected and the moduli space admits a much simpler
description.  Indeed, well separated D0-branes are  
characterized by the complete diagonalization of all the
adjoint 0-0 fields corresponding to the space-time positions, i.e.\ we
consider $Z, \tZ$ diagonal generic matrices, which spontaneously break
$U(k)$ to the abelian subgroup $U(1)^k$. It is intuitively clear that
this produces just $k$ copies of the single instanton moduli space.   
In this subspace, the $4k$ position degrees of freedom $z_i , \tz_i$ are
completely decoupled from the rest of the variables.  The
 hypermultiplets  $D$-equations collapse to   
\eqn\hdeq{
\left( \phi_h \phi_h^{\dagger} - \tphi_h^{\dagger} 
\tphi_h \right)^m_m=0,}
and the $F$-equations, for diagonal $W$ and $W'$ fields:
\eqn\hfeq{
\eqalign{ (w_m - w'_a)(\phi_h)^m_a =& 0 \cr
          (w_m - w'_a)(\tphi_h)^a_m =& 0 \cr
          (\phi_h)^m_a (\tphi_h)^a_{m} = & 0}}
where  $m =1, \ldots , k $ is an index labeling the  D0-brane  corresponding
to each $U(1)$ factor,   and $a,b,\ldots$ 
are the flavour D4-indices.
The interesting branch of solutions is the one with maximal Higgsing of
the hypermultiplets, with $w_m = w'_a=0$, all $m,a$ (all D0 and D4-branes 
on top of each other in the $(x^4, x^5)$ plane).  
Subtracting the $2k$ $F$-equations
and the $k$ $D$-equations from the total $4kn_c$ real components in
the scalars $\phi_h, \tphi_h$, and further dividing by the $U(1)^k$
gauge symmetry, we find a total of $4kn_c -2k-k-k = 4k(n_c -1)$ real
parameters. Adding the $4k$ degrees of freedom coming from the eigenvalues
of $Z$ and $\tZ$, representing the translation moduli in the physical
dimensions, we end up with the correct $4kn_c$ dimensional bosonic moduli 
space of YM instantons.

The previous parametrization is appropriate for the dilute instanton limit, 
in which the non-abelian structure of $U(k)$ is not really probed. Indeed,
when the instantons, or D0-branes ``overlap", the off-diagonal 
entries in the $Z, \tZ$ fields become important 
and eqs.\ \hdeq\ and \hfeq\ are
modified by terms which mix the adjoint superfields with the 
hypermultiplets. In any case, in the Higgs branch relevant to our discussion,
$Z$ and $\tZ$ contribute $4k$ degrees of freedom and the hypermultiplets
$4k(n_c -1)$, describing as before the correct $4kn_c$ dimensional 
bosonic moduli space of YM instantons.

\newsec{D-brane construction of exact instantons in SYM}

In this section we adapt the previous construction to the more
detailed realizations of SYM theories in four dimensions with
different amounts of supersymmetry, starting with the maximal
one, $N=4$ SYM.  
 
\subsec{N=4 SYM}

Since the world-volume theory of the D4-branes already has
 $N=4$ $SU(n_c)$ 
SYM theory in five dimensions, we simply have to compactify the
D4-D0 bound states to four dimensions, without any further breaking
of supersymmetry. Consider wrapping
both the D4 and D0-branes around a circle of length $L_6$ in the $x_6$ 
direction.\foot{Recall that this corresponds to the euclidean time
of the D0-branes.}  On
the relevant bosonic moduli space of real dimension $ 2d= 4kn_c$, the 
D0-branes world-line has a set  of $d$ free superfields $(\xi_s, \eta_s)$, 
in four-dimensional notation, i.e.\ we have an action
\eqn\accl{
S_{D0} =k M_{D0} \int_0^{L_6} d\tau - ik\theta + {1\over g_s} 
\int_0^{L_6} d\tau
\sum_{s=1}^{d} \left( |{\dot \xi}_s |^2 + i
 \,{\overline \eta}_s {\dot \eta}_s 
\right) + {\rm massive\ .}} 
Each  Weyl fermion $\eta_s$ has two complex or 
four real components off-shell so that, 
when the
one-instanton contribution to the path-integral is dominated by the
classical static trajectory ${\dot\xi}_s = {\dot \eta}_s =0$,  we
get a factor
\eqn\instf
{Z_{\rm inst} = \int \prod_s d{\overline\eta}_s d\eta_s \, d\mu(\xi_s)\,
e^{-kM_{D0} L_6 + ik \theta
} = 0^{4d} \, {\rm Vol}({\CM}) \, e^{-{8\pi^2 k \over g^2} 
+ik\theta} }    
where we defined $M_{D0} L_6\equiv 8\pi^2 / g^2$. 
In this expression, the bosonic measure $d\mu(\xi)$
for collective coordinates contains zero-mode Jacobians  which are most
easily derived in the field theoretical framework. The corresponding
integral gives the formally divergent volume of moduli space, 
${\rm Vol} (\CM )$. The symbol $0^{4d}$
denotes the number of fermionic zero modes of the instanton, to be 
saturated by fermion sources. We see clearly that the number of zero
modes corresponds to the total number of independent anti-commuting 
collective coordinates, which in turn coincide with the total number
of off-shell fermion components on the D0-brane world-line.

 The first important check of expression
\instf\ is this number of zero modes $4d = 8kn_c$ which is indeed 
the right one for $SU(n_c)$ SYM in the $k$-instanton sector. In field theory, 
each gluino system gives $2kn_c$ zero modes, and we have four independent
gluino systems in $N=4$ SYM. In the D0-D4 branes construction,
the $8k$ zero modes associated to the
$Z, \tZ$ superfields are interpreted in the four dimensional SYM theory
as supersymmetry zero modes. On the other hand, $8k$ out of the $8k(n_c-1)$
zero modes associated
to the hypermultiplets $H, \tH$ are interpreted in space time as
superconformal zero modes, while the remaining  $8k(n_c -2)$ zero modes
do not have an obvious symmetry interpretation, since they arise
as 't Hooft zero modes associated to doublets with respect to the
$SU(2)$ subgroup of the gauge group where the instanton sits.    

The second check of \instf\ is given by the fit of the Yang-Mills coupling
to the mass of the D0-brane, $M_{D0} L_6 \equiv 8\pi^2 /g^2$, needed
to reproduce the standard instanton factor.
 Dimensional reduction from the D4-brane Born-Infeld
action leads to
\eqn\dimrdf
{{\tau_4 } \int_{0}^{L_6} d\tau \,\Tr\,\sqrt{{\rm det}(
1+2\pi\alpha' F)} \rightarrow -{\tau_4 \over 4} L_6 (2\pi\alpha')^2
 \,\Tr\,F^2 \equiv
-{1\over 4g^2_{YM}} \, \Tr\, F^2.}
Using $\tau_p^{-1} = g_s \sqrt{\alpha'} (2\pi \sqrt{\alpha'})^p$ and
$M_{D0} = \tau_0$ we find agreement with \instf, i.e.\ $g^2=g^2_{YM}$.    
This is an important point, since it implies that the length of the
D0-branes 
world-line to be considered is really $L_6$, a statement which will
become non-trivial in the next section, when considering the
case of $N=2$ and $N=1$ SYM theories. 

\subsec{N=2 and N=1 SYM}
More interesting is the D-brane construction of instantons with
reduced supersymmetry in the four physical dimensions. Using the
constructions of refs.\ \rhw\ and \regk, we compactify the five-dimensional 
theory to four dimensions on the D4-brane world-volume by means of a 
Kaluza-Klein reduction on a segment of length $L_6$ in the $x^6$ direction, 
bounded by NS five-branes with world-volume in
the $(x^0,x^1,x^2,x^3,x^4, x^5)$ directions. By a suitable complex rotation 
of the $(x^4,x^5)$ plane into the $(x^8, x^9)$ plane
of one of the NS-branes \refs\rbarb, one breaks further
the $N=2$ supersymmetry to $N=1$. In such constructions,
the role of the NS-branes is simply to project out certain degrees of
freedom, and  the appropriate
supersymmetries. They   do not  
add any extra degrees of freedom, unless special classes of phase
transitions are considered.
In particular, the fermions propagating in the world-volume of the
NS-branes are set to their vacuum values. Among them, we have the
six-dimensional Goldstone fermions associated to the supercharges broken
by the NS-brane. Therefore, in constructing the relevant Yang-Mills
instantons using D0-branes, it is natural to project out those fermion
zero modes which overlap with the above-mentioned six-dimensional 
Goldstone fermions. This makes sense because we consider constant
fermion configurations, ${\dot \eta}=0$, on the
D0-brane world-line, which extends between the NS branes, intersecting 
them at the end-points.

 Recalling \dimrdf, we see that the D0-brane
world-line has length $L_6$ in order to fit the bare Yang--Mills
instanton action, and therefore it extends only once between the
NS-branes. This world-line touching the NS-branes at the end-points
must be considered as topologically stable, unlike the case of
free D0-branes or the $N=4$ setting without NS-branes, where
a finite D0-action contribution requires wrapping the world-line
around some non-contractible circle in the target space. This 
distinction between free D0-branes and bound D0-branes will
reappear in our comments  on the relation with the M-Theory 
lifting in the last section.

On the other hand, it is important that no extra constraints are 
imposed by the NS-branes  on the bosonic collective 
coordinates $\xi_s$, whose measure
is completely determined in \instf\ in the $N=4$ theory, up to
Jacobian factors which are related to zero modes. In the absence of
a detailed understanding of the microscopic couplings between D-branes
and NS-branes, we have to rely on consistency checks with the field
theory picture and geometric arguments. For example, the projection
of certain bosonic degrees of freedom on the world-volume of the
D4-branes follows from the global geometry once we assume the
rigidity of the NS-branes. Alternatively, the bosonic projections
follow from the fermionic ones by supersymmetry. In our case, the brane
geometry projects out the scalars in the 0-0 sector, except those
representing the motion of the D0-brane in the physical space-time
$(x^0, x^1, x^2, x^3)$, but there
is no geometric constraint on the hypermultiplet degrees of freedom
of the 0-4 and 4-0 sectors, and therefore no obvious geometric 
constraint imposed by the NS-branes  on the associated collective 
coordinates $\xi_s$. The intersection of the
D0-brane world-line with the NS branes occurs at isolated points
in space-time and, as a result, there is no hamiltonian representation
of bosonic and fermionic degrees of freedom on the instanton ``world-volume'',
and therefore no statement of equality between fermionic and bosonic
``states", as it would be required by supersymmetry in non-vanishing
dimensions. Exactly the same mechanism is at work in the standard 
instanton superfield formalism \refs{\ritep,\rrev,\rdorey}. 

In order to see how the fermion projections work, consider
the case of two NS-branes and a set of $n_c$ D4-branes and $k$ D0-brane 
world-lines suspended between them in the $x^6$ direction. Denote by 
NS' the second NS-brane, which is parallel to the first NS-brane but 
translated in the $x^6$ direction by $L_6$ in the
$N=2$ configuration, and rotated into the $(x^8, x^9)$ directions
for the $N=1$ configurations.
We have seen that $8k$ of the fermionic collective coordinates in
the bulk come from the fermionic superpartners of the superfields
$Z, \tZ$, representing translations in the space-time $(x^0, x^1, x^2, x^3)$
directions. In the D-brane construction, they arise in the 0-0 sector, 
and can be interpreted as part of the Goldstone fermions of the broken 
supersymmetries by the D0-brane world-line.
An isolated D0-brane breaks 16 of the total 32 supercharges of the
type-IIA theory, and accordingly we find 16 Goldstone fermions on the
world-line. These fermions where assembled in the first section into
four four-dimensional Weyl fermions:
$\psi_{z},
 \psi_{\tilde z}, \psi_w , \psi_v$. 
As we have stressed in different occasions, only half of them, namely
$\psi_z, \psi_{\tilde z}$ survive the Higgs mechanism on the branch of
D0-D4 branes bound states.  
 
In general, in the present type-IIA context, the projector over the unbroken
supersymmetries on a D-brane world-volume is best represented in
M-Theory language  as
\eqn\prj{P_p = {1\over 2} (1+i \Gamma (\Sigma_{p+1}))}
where $\Gamma (\Sigma_{p+1})$ is the product of Dirac matrices in the
world-volume directions, with the understanding that $p$-branes whose
M-Theory description involves a wrapped or boosted M-brane in the    
eleventh direction include a factor of $\Gamma_{11}$ \refs\rangles.
 The factor of
$i$ in \prj\ stands for the euclidean rotation of $\Gamma_0$, as 
appropriate for the instanton discussion. So, in terms of a 32-dimensional
spinor $\Psi_{32}$, the 16 Goldstone fermions carried by each D0-brane
are given by $(1-P_{D0})\Psi_{32}$. 
Now, according to  the
prescription above, we should keep only those zero modes with no overlap
with the Goldstone fermions on the NS-brane world-volume. The remaining zero
modes from the 0-0 sector of a general D0-brane  are then 
\eqn\remz{ \eta_{z, {\tilde z}, v, w}
 = (1-P_{D0}) P_{NS} P_{NS'} \Psi_{32}\ .}
Using the explicit form of the projectors, one easily sees that each
NS-brane divides by half the number of zero modes, so that we have 8
zero modes in the $N=2$ configuration, and 4 zero modes in the $N=1$
configuration.

 The previous considerations apply to the situation of a free D0-brane.   
In the case of a D0-D4 branes bound state we have two modifications. First,
as stated before, on the Higgs branch of bound states the fermions
$\psi_v , \psi_w$ are lifted, and the number of zero modes \remz\ coming
from the 0-0 sector is again divided by half. Second, we have a new
set  of degrees of freedom, namely the $\psi_h, \tpsi_h$ from the
0-4 and 4-0 sectors. Those are constructed   
 from the zero modes of the world-sheet fermions $\psi_0^{\mu}$ for 
$\mu=4,5,6,7,8,9$ (i.e.\ in the NN $+$ DD directions of the D0-D4 brane
system), in the Ramond sector.  
Indeed, there are $\half 2^{6/2} = 4$ field components (off shell), after
GSO projection, in the 0-4 sector, that is a Weyl fermion. In terms of the
complex combinations $\Gamma_u = \half (\Gamma_6 +i\Gamma_7)$, $
\Gamma_w = \half (\Gamma_4 +i\Gamma_5)$, $\Gamma_v = \half (\Gamma_8 +
i\Gamma_9)$, the chirality operator reads
\eqn\gsop
{\Gamma_{11} = i\Gamma_0 \Gamma_1 \cdots \Gamma_9 =-(1-2N_z)(1-2N_{\tilde z})
(1-2N_u)(1-2N_v)(1-2N_w), }
with $N_v \equiv \Gamma_v^{\dagger} \Gamma_v $, etc., the fermion occupation
numbers taking values 0 or 1. Acting on the $2^{6/2} + 2^{6/2}$ states in
0-4 and 4-0 sectors, built as polynomials $P(\Gamma_u^{\dagger},
\Gamma_v^{\dagger}, \Gamma_w^{\dagger} ) |0\ket$, the GSO projection
reduces to $-(1-2N_u)(1-2N_v)(1-2N_w) =+1$. On the same set, the 
projection imposed by the NS brane is $+1 = i\Gamma_{NS} = i\Gamma_0 
\Gamma_1 \cdots \Gamma_4 \Gamma_5 = -(1-2N_w)$. So, on this subspace
the NS projector is $P_{NS} = N_w$, and the analogous
projector for the rotated NS'-brane is $P_{NS'} = N_v$.  

We see that, in all cases, the effect of each NS-brane 
is to divide by 2 all the bulk zero-modes democratically.  
This amounts to a total factor of $1/4$ when the NS' brane is a rotation
of the NS-brane into the $(x^8, x^9)$ plane. So,
if $N$ denotes the amount of space-time
supersymmetry of the brane configuration, the resulting instanton
leaves $2N$ supercharges unbroken, out of the total $4N$ which are 
linearly realized in the vacuum of the space-time theory. 
The total number of fermionic zero-modes is given by:
\eqn\totz{{\cal N}_{f.z.m.} = {N\over 4} \cdot 4d 
  = {N\over 4}\cdot  8 k + {N\over 4}\cdot 8k(n_c-1) = 2Nkn_c,  } 
the correct number, where
we have already subtracted the massive degrees of freedom in the
 Higgs branch which defines the instanton moduli space,
and we have split the zero modes between those corresponding to
the $2N$ broken supercharges,  and those of a fixed instanton, coming from
the hypermultiplet degrees of freedom.  

Notice that in the preceding sections, it was convenient to adopt
the notation in terms of Weyl fermions with respect to the $SO(1,3)$
of the $(x^6, x^7, x^8, x^9)$ plane. However, the presence
of the NS or NS'-branes breaks this group, and the fermion components
left after projection need not fullfill representations of such group.

\newsec{Constrained instantons and the Coulomb branch}

Up to now we have considered the case of instantons in super Yang-Mills 
theory and showed that it can be described by a system of $k$ D0-branes 
bound to $n_c$ D4-branes. In particular this means that the D0-branes and
the D4-branes are all located at the same point in the 
$(x^4,x^5,x^7,x^8,x^9)$ plane. In this section we will show that the
geometrical operation of moving apart the D4-branes corresponds, as 
intuition would tell, to describing constrained instantons
 \refs{\rtof,\raffl,\rfy}. 
We postpone the introduction of space-time matter, i.e.\ D6-branes, 
and its complications to the next section. Thus in this section we will
study the Coulomb phase of $N=2$ SYM models. 

Let us start by considering what happens in moving apart the D0-branes. 
For example, separating one D0-brane from the 
others in the $(x^4,x^5)$ directions, just means
that one is considering a bundle of $(k-1)$ instantons. In other words,
only when the D0-branes are bound  to the D4-branes  
(in the $(x^4,x^5,x^7,x^8,x^9)$ plane) they can describe exact instantons. 
It obviously follows that, in moving apart a D4-brane from the others,
we will leave the D0-branes bound to the $(n_c-1)$ remaining D4-branes, and
that we can only speak of exact instantons on the corresponding $SU(n_c-1)$
subgroup. We can, however, still consider approximate (``constrained'')
instantons of the full $SU(n_c)$ group, by moving a little apart, 
one by one, the D4-branes while keeping the D0-branes bound 
to at least two D4-branes.\foot{Indeed in field theory we need at 
least $SU(2)$ to be able to build YM instantons. Analogously, the D-brane 
construction collapses if $n_c<2$, because there is no $N=2$ Higgs
branch for less that two flavours.} 

In the following we shall concentrate on the single instanton sector,
i.e.\ only one D0-brane, and return at the end of the section to the
more involved case of multi-instanton interactions. For a single D0-brane
only the relative position with respect to the D4-branes matters, and
therefore we may fix the D0-brane at $w=0$, and consider the effect of
turning on the $W'$ field, whose diagonal configurations correspond to
well defined D4-brane positions. 

The relevant coupling in the D0-brane world-line theory
 which is sensitive to  the D4-branes motion is (see  
eq.\ \ocsup)
\eqn\lfour{
\CL_{04-44-40} = {\sqrt{2}\over g_s}
\int d^2\theta \,\Phi_{04} \Phi_{44} \Phi_{40}
+ {\rm h.c.}}
This coupling is simply a mass term (the mass matrix given by
$\phi_{44} \sim W'$ which describes the position of the D4-branes in the 
$(x^4,x^5)$ plane) for the instanton moduli (in the notation of
\ocsup\ $\Phi_{04} \sim \tH$ and $\Phi_{40} \sim H$). When considered in the
Coulomb branch of the $N=2$ configurations, it yields the familiar
action of a constrained instanton:
\eqn\Sconstr{
S_{\rm constr} = {1\over g^2} (8\pi^2 + \rho^2 v^2 ) \ ,
}
with $v \sim \bra \phi_{44} \ket$ and $\rho^2 \sim \phi_{04}\phi_{40}$.
At the same time, it lifts the fermion zero modes through the
pairings $v \psi_{04}\psi_{40}$. For the fermions, the lifting of
zero modes occurs homogeneously in the bulk of the D0-brane world-line, 
and therefore it is obviously compatible with the projections imposed 
by the NS-branes at the boundaries. 

In general, an homogeneous lifting pattern of this sort, appropriate
to describe the Coulomb phase of $N=2$ models, gives a mass of order
$v$ to $d_c$ complex 
bosonic collective coordinates, and similarly, it lifts
$4d_c$  real fermionic coordinates in the bulk, i.e.\ it induces
 a term in
the bulk action of the form
\eqn\lba{
\delta\CL \sim v^2 \sum_{s=1}^{d_c} |\xi_s |^2 + v\sum_{s=1}^{d_c} 
{\overline \eta}_s \eta_s \ ,  
}
where we have rescaled the bare Yang-Mills coupling $g^2 \sim g_s /L_6$ 
into the collective coordinates $\xi_s$ and $\eta_s$.   
Taking into account the projection at the boundaries of the world-line 
due to the NS and NS'-branes, we have a partition function of the form
\eqn\Zconstr{
Z_{\rm inst}
 \sim 0^{N(d-d_c)}\,\, {\rm Vol}_c (\CM) \,\,v^{Nd_c \over 2}\,\,  
e^{-{8\pi^2 \over g^2} +i \theta } 
}
where ${\rm Vol}_c (\CM)$ represents the bosonic ``volume" of the
moduli space, regulated with the gaussian term ${\rm exp}(-v^2 |\xi|^2)$,
and the term $v^{Nd_c /2} $ comes from the fermionic integrals.\foot{
Recall that $N$ is the number of four-dimensional
 supersymmetries in the space-time
configuration, and $d =2n_c$ is the complex dimension of the 
bosonic instanton moduli space.} 
The complete power of $v$ in the full instanton
measure depends on the result of the bosonic integrals, of course.
In general, we must view the couplings like \lfour\  as small
perturbations over the instanton moduli space described in the 
introduction. This amounts to the requirement that the dimensionless
effective expansion parameter $\rho v$ be small.

The important remark to make is that all zero modes due to the 
conformal symmetry and embedding angles, namely, all zero modes
encoded in $\Phi_{04}, \Phi_{40}$, can be lifted in this way on the
Coulomb branch of the $N=2$ theory, by switching on $\Phi_{44}$ (i.e.\      
splitting the D4-branes in the $(x^4, x^5)$ plane). This amounts
to lift generically 
$2d_c = 4(n_c -1)$ bosonic and $4(n_c -1)$ fermionic zero modes.
The remaining zero modes, $4$ bosonic and $4$ fermionic, correspond
to the superfields from the 0-0 sector $Z$ and $\tZ$, which remain
uncoupled and therefore are not lifted. Since these collective 
coordinates are
associated to the translations in the physical four dimensions,
these fermion zero modes are naturally interpreted as
due to the broken supersymmetries of the instanton, understood
as a classical solution in the physical four-dimensional
theory on the fourbranes.

We now briefly sketch the situation in the case of multi-instanton 
configurations. In the  limit of dilute instantons, the lifting 
pattern of the  zero-modes  is just given
by $k$ copies of the single instanton case. Instead, beyond the
dilute approximation,  new terms appear in the effective lagrangian,  
due to the non-vanishing of the off-diagonal entries of the $Z, \tZ$ fields. 
For example, in the parametrization given in ref.\ \refs\rwihiggs\ of the
Higgs branch of the effective theory of the D0-branes, one can 
exhibit particular configurations with diagonal $Z$, but
non-trivial off-diagonal entries of the $\tZ$ field,  given by
\eqn\ofd{ \tZ^m_n\ \sim\ -{(\phi_h)^m_a (\tphi_h)^a_n \over z_m - z_n}} 
as a function of the eigenvalues $z_n$ of the $Z$ fields, 
and the hypermultiplets' moduli. Now the term $|[W,\tZ]|^2 $ of the 
effective potential vanishes unless at least 
one of the two fields has non-vanishing
off-diagonal entries. Thus in presence of the non-trivial off-diagonal 
entries of the $\tZ$ field given by eq.\ \ofd, and turning on a diagonal $W$
field, representing definite separations of the D0-branes in the 
$(x^4, x^5)$ plane, we find positive energy contributions to the 
effective potential of the form  
\eqn\runa{ {|w_m-w_n|^2\over |z_m-z_n|^2} (\phi_h)^m_a (\tphi_h)^a_n
(\phi_h^*)^n_b (\tphi_h^*)^b_m}
which describe a runaway potential for the eigenvalues $z_n$, and
an analogous lifting term for the superpartners. Notice that
this potential depends only on the relative separations 
$z_m-z_n$, so that the overall 
center of mass of all D0-branes is not lifted by these interactions,  
and furthermore \runa\ vanishes  
in the limit of instantons at very large distances, i.e.\ the dilute 
limit. This is in agreement with the known results in field
theory (see for example ref.\ \refs\rdorey) where it has been shown 
that on the Coulomb branch there remain only 4 bosonic and 4 fermionic
zero modes, associated with the center of mass position of the instantons
and with the broken supersymmetries, respectively. 

\newsec{Space-time flavour and the Higgs branch}

In the D-brane formulation of super-QCD (SQCD),
 the matter (space-time) flavour 
is described by $n_f$ D6-branes with world-volume in the 
$(x^0,x^1,x^2,x^3,x^7,x^8,x^9)$ directions. We know from field
theory that the main effect of the presence of matter
 on the instantons is the appearance
of new fermionic zero modes, as well as new patterns of zero
mode lifting when turning on the Higgs branch of the field theory.
In this section, we will start by describing the flavour zero
modes of the instanton in terms of the local dynamics of the D0-branes and
the D6-branes.   An important aspect of this dynamics, and a source
of  many subtleties, is the non-standard geometrical arrangement of the
D0-D6 branes system. In our construction, the D0-branes are extended in
euclidean time in the $x^6$ direction, while the D6-branes are
completely localized on this axis. Therefore, D0 and D6-branes do
not share any propagating dimension, and are purely instantonic
relative to each other. This changes the standard statement that
the D0-D6 branes system is not supersymmetric. Here the relevant parameter
$\nu = d_{ND}+d_{DN}$, the sum of ND and DN directions in the boundary
conditions of 0-6 and 6-0 strings, takes the value $\nu =8$ and we
have some unbroken supersymmetry. 

The local spectrum on the 0-6 and 6-0 strings is similar to the
standard $\nu=8$ spectrum at the intersection between a D1-brane
and a D9-brane, as explained for example in \rpol. We can transform
the D0-D6 brane system at hand into a D1-D9 brane pair by means of a  
T-duality transformation along the $(x^4, x^5,x^6)$ directions and 
considering $x^4$ as the (euclidean) time. 
In the NS sector the Casimir energy of
0-6 strings is $-1/2 + \nu /8 = +1/2$ and therefore there
are no bosons at all made from 0-6 strings. In the Ramond
sector the worldsheet fermions $\psi^{\mu}$ in the ND or DN directions
have half-integer moding, and so the corresponding vacuum is    
non-degenerate. We are left with the two fermionic zero modes
in the two DD (or NN under T-duality) directions. This
leads to two states, one of which is projected out by the GSO
projection. The physical state condition, or Dirac equation
$G_0 = p_{\mu} \psi^{\mu} =0$, which imposes a holomorphic
constraint on the D1-brane excitations of the D1-D9 brane system, here
degenerates due to the instantonic nature of the D0-D6 intersection.  
 In our case, where
the D0 and the D6-branes do not share any propagating dimensions, we
have no physical state condition. In all, adding the 0-6 and
6-0 sectors, we would have a total of $2kn_f$  fermionic
collective coordinates, the correct number we expect from field
theory considerations. Due to the absence of a physical state condition,
we have to think of these localized states as effectively off-shell, and
therefore, we must have $2kn_f$ holomorphic Grassmannian integrals in the 
instanton measure, taking the form 
\eqn\flavm{d\mu_{\rm flavour} =
 \prod_{j=1}^{n_f} d\chi^*_j d\chi_j = d\chi_{60} d\chi_{06}.}     
The two states at each D0-D6 branes intersection can be assembled into
a complex field $\chi =\chi_{06}$, with the orientation reversed
component $\chi_{60}$ representing the conjugated field $\chi^*$. 
In this way we may realize the $U(1)$ gauge symmetry of the D6-brane
as a global symmetry on the D0-D6 branes intersection.\foot{Notice that,
in this picture, the full $U(n_f)$
global flavour symmetry is hard to visualize explicitly,
 since the D6-branes sit at different
``instants" of time from the point of view of the D0-brane 
world-line.}

Now we can consider the full system adding the D4-branes and 
the NS-branes. By adding the D4-branes the space-time theory
becomes SQCD with $n_f$ flavours 
and we expect that in the world-line theory of the
D0-branes there will appear induced couplings due to the 4-4 and 4-6 
strings.  As before the NS and NS'-branes
give us $N=2$ or $N=1$ SQCD. The only new feature is the effect of 
the projections induced by the NS and NS'-branes on the 0-6 strings. 
The claim is that there is no new projection and thus the number of
zero modes due to the 0-6 strings does not change with respect to the
analysis in the previous paragraph.  This can be understood as follows. 
Recall that the D6-branes are completely instantonic with respect to
the zero branes and that in the $x^6$ space they are generically located
between the NS and NS'-branes. Thus 0-6 strings are localized at the 
intersection of the D6-branes with the D0-branes world-line.
 For this reason they
do not feel the presence of the NS and NS'-branes (unless of course
the D6-brane is on top of the NS or NS'-brane, but this is a non generic  
configuration that we will not consider). Notice that the situation
is quite different for the 0-4 strings, since the D0-branes are bound to 
the D4-branes in the $x^6$ direction, the 0-4 strings are not localized
and thus feel the projections due to the NS and NS'-branes at the 
boundaries. The same of course happens for the 0-0 strings. 
The independence of the 0-6, 6-0 modes upon the boundary projection
is in beautiful correspondence with the independence of the quark
instanton sector upon the amount of supersymmetry in SQCD.

After our description of the exact instanton in the SQCD at the
origin of moduli space, we can discuss the patterns of zero-mode
lifting when exploring the moduli space of the space-time theory.
In the brane picture this corresponds to motions of the background
branes, and the corresponding liftings are understood in terms of
induced couplings in the D0-branes world-volume.
As before, we will concentrate mainly on the single instanton sector
(or the dilute instanton case).

First of all, there is the trivial case of adding a mass to the
quarks in space-time, or moving the D6-branes away from the D4-branes
in the $(x^4,x^5)$ plane.
This can be described on the D0-branes world-line as a local coupling of
the $\chi_{06}, \chi_{60}$ variables to the scalar $\phi_{66}$
taking a non-vanishing expectation value under relative D4-D6 branes
displacement:
\eqn\zmm{\CL_{06-66-60} \sim \chi_{06}\, \phi_{66}\, \chi_{60} \,
\delta(\tau - \tau_{D6}).}
As expected, this coupling lifts the associated flavour zero
modes of the instanton, leaving all other fermionic and bosonic
collective coordinates intact.  

A more complicated pattern appears when exploring the Higgs branch
of the space-time theory, by switching on the expectation values of  
the squark fields $\bra \phi_{46} \ket, \bra \phi_{64} \ket \neq 0$. 
In the brane picture, D4-branes corresponding to the Higgssed color
indices are broken on the D6-branes and lifted into the
 $(x^7, x^8, x^9)$   
directions in the $N=2$ configurations, or into the $(x^8, x^9)$ plane
in the case of the $N=1$ setting.  Now, the corresponding instanton
coordinates, built from some components of the $\Phi_{04}, \Phi_{40}$
superfields, get masses due to the finite stretching of the 0-4 and
4-0 strings.  This stretching occurs in the bulk of the D0-brane
world-line, along the lifted portion of the D4-brane, and not only
at the location of the D6-branes. Therefore, the bulk mass
of the lifted $\Phi_{04}, \Phi_{40}$ fields is due to a coupling
which can be written with no reference to the D6-branes. It simply
corresponds to an expectation value of the  $\Phi_{44}$
fields in the appropriate directions.  Now, in the notation adopted
in section 1, in order to write the D0-brane world-line supersymmetry
in four-dimensional superspace notation, we represented the
D0-D4 fields as T-duals of a D3-D7 brane system whose intersection lied
in the ``space-time" $(x^6, x^7, x^8, x^9)$. In this notation, a
motion of the D4-brane in the $(x^8, x^9)$ plane is represented
as a background gauge field $V'_{44} = A'_{8} + iA'_{9}$ on the
D7-brane world-volume. Going back to the D0-brane world-volume
via dimensional reduction, we have to consider a non-zero
expectation value of the ``gauge field" $V'_{44}$, coupling
to the ``flavour indices" of $\Phi_{44}$. So, from this point of view,
the appropriate masses come now from D-terms rather than holomorphic 
couplings on the D0-branes world-line.  Taking 
$\bra \phi_{v'} \ket \sim \bra \phi_{46} \ket \sim \bra \phi_{64} \ket \sim v$,
we have induced masses of the form  \lba\ for the bosonic and fermionic
components of the $\Phi_{04}, \Phi_{40}$ superfields, containing
the collective coordinates of a fixed instanton.

On the other hand, the  $2n_f$ flavour collective coordinates
 found above, $\chi_{06}, \chi_{60}$,  are  localized at each D0-D6 
intersection, and must be lifted by means of a local coupling 
involving all three types of branes touching locally at that
point, including the D4-branes. The reason is that, as we know from
the field theory analysis, the flavour zero modes are lifted by
pairing them with the rest of the fermionic zero modes of the 
instanton. In the D-brane setting, we must pair up the flavour
collective coordinates with the fermion zero modes from the
0-4 and 4-0 sectors.  Now, the most general coupling respecting
the orientation reversal symmetry of the string theory,  
hermiticity, and the $U(1)$ global symmetry of the D6-branes, 
is given by   
\eqn\lsix{
\CL_{06-64-40} \sim \left[\chi_{06}\Big(\phi_{64}\psi_{40} + \phi_{46}^* 
\psi_{04}^*\Big) + \chi_{60} \Big( \psi_{04}\phi_{46} + \psi_{40}^* 
\phi_{64}^* \Big) \right] \,\delta(\tau- \tau_{D6}) \ ,  
}
where we have used the fact that the complex conjugate of a 0-4
field, $\psi_{04}^*$, transforms like a 4-0 field. 

In order to exhibit in more detail the pattern of pairings, it is
convenient to consider in turn the two cases with $N=2$ and $N=1$
space-time supersymmetry. In the first case, the NS projection leaves
$4n_c$ holomorphic fermionic  collective coordinates
of the 0-4 and 4-0 strings, prior to any
Higgsing on the instanton moduli. We can denote them as  
$\psi_a , \tpsi_a^*$,
transforming as $\psi_{04}$, and 
$ \tpsi_a, \psi_a^*$, transforming as $\psi_{40}$. The part of the
$N=2$ Higgs branch accessible to the D-brane representation is the set
of  non-baryonic Higgs branches,  
which,  according to ref.\ \rdeff,  can be parametrized,
up to $U(n_c)\times U(n_f)$ transformations, by    
\eqn\nbhiggs{
(\phi_{64})^a_j = v_j \delta^a_j\ ,  \qquad\qquad (\phi_{46})^j_a
 = v_{j-[n_f/2]} \delta^{a+[n_f/2]}_j 
}
where $a=1,...,n_c$,  $j=1,...,n_f$ and at most $r\leq[n_f/2]$ of the
$v_j$ are non zero. Then, the general  coupling \lsix\  takes
the form
\eqn\sixpair{
\CL_{\tau =\tau_{D6}} \sim \sum_{j=1}^{r} v_j \Big(\chi_j \tpsi_j + 
\chi_j^* \tpsi_j^* \Big) + \sum_{j=[n_f/2]+1}^{r+[n_f/2]} v_{j-[n_f/2]} 
\Big( \chi_j \psi_{j-[n_f/2]}^* + \chi_j^* \psi_{j-[n_f/2]} \Big) 
}
and we see clearly the geometric pattern of lifting characteristic of
the $N=2$ configurations: two D6-branes are necessary to lift one D4-brane, 
and also the zero mode lifting pattern is two to one, in the sense
that two fermion components associated to the same D4-brane pair up
with two flavour components associated to different D6-branes, those
on which the D4-brane is sliding.  

In the case of the $N=1$ configuration, the NS-NS' projection leaves
only $2n_c$ fermionic components of the 0-4 and 4-0 strings, 
which we can denote by  $\psi_a, \psi^*_a$.    
 Also, the Higgs constraints are different, and the Higgs
phase is parametrized by $\phi_{46} = \phi_{64} = {\rm diag}(v)$.
Therefore, the general coupling \lsix\ reduces now  to
\eqn\sixone{
\CL_{\tau=\tau_{D6}} \sim \sum_{j=1}^{r} v_j \Big(\chi_j \psi_j + 
 \chi_j^* \psi_j^* \Big) \ .
}
Now the pairing is one to one, as only one D6-brane is needed
to lift each D4-brane.    
 
The couplings \sixpair\ and \sixone\ must be considered together
with the induced masses in the bulk 
for the $\psi_{04}, \psi_{40}$ fields, coming from the D-term 
couplings to the $V'$ field. In addition, we have the masses
induced by the Higgs effect on the instanton moduli space, i.e.\
the equations \hdeq\ and \hfeq, which must be regarded as
dominant, according to the prescription stated in section 3, that
all couplings must be considered as  perturbations of  the instanton
moduli space, i.e.\ $\rho v << 1$. Thus, we have a complicated pattern
of mixings between the fermionic zero modes on the Higgs branch. 

In particular, in the case of $N=2$ supersymmetry and $n_f=2n_c-2$ flavours 
on the non-baryonic branch, one can lift all flavour zero modes and 
$4(n_c-1)$ fermionic zero modes of the fixed instanton. Only the $4$ 
fermionic zero modes associated to the $Z$ and $\widetilde{Z}$ fields remain,
since these fields are spectators and are not lifted on all the branches 
of the model. This of course agrees with the field theory results. 

Similarly, in the case of $N=1$ supersymmetry with $n_f=n_c-1$ flavours, 
on the Higgs branch one can lift all flavour zero modes and $2(n_c-1)$ 
fermionic zero modes of the fixed instanton. The $2$ remaining zero modes
coming from the $Z$ and $\widetilde{Z}$ fields,  are those
 related to the appearence of the Affleck-Dine-Seiberg 
super-potential \refs\rads. 

Thus in the case of $N=1$ supersymmetry with $n_f=(n_c-1)$ D6-branes, 
the presence of a D0-brane bound to the D4-branes gives rise, on the Higgs 
branch, to an effective repulsive interaction between the D4-branes. 
This effect is a particular case of the ``quantum rules" proposed in
 ref.\ \regkrs\ for even more general configurations 
with $N=1$ supersymmetry and an arbitrary number of D4-branes and
D6-branes. 
The D0-branes, which act as the field theory instantons in these models,
give rise to this ``quantum rule'' only in the $n_f=n_c-1$ case. 
In fact, a more complete motivation of such quantum rules in terms
of instantons requires studying the system compactified on a circle,
and exploiting T-duality \refs{\rberk,\regkrs}. 
Under a T-duality transformation, D0-branes
are converted into D1-branes of type-IIB string theory, and the
associated instanton configurations look like dimensionally reduced
monopoles of the theory with an additional adjoint scalar. The superpotentials
on the Coulomb branch of this scalar are saturated by three-dimensional
instantons for any number of flavours. This  ``fractionalization"
property of instantons upon compactification  is a  
well known  phenomenon \refs{\rgomez,\ryi,\rrut}.

\newsec{Discussion}

In this paper we have studied the microscopic model of 
 instantons in $SU(n_c)$ SYM theories
as D0-branes bound to D4-branes in the Type-IIA D-brane description of the
space-time theory. In particular, we have traced the phenomenon of
zero-mode lifting of constrained (approximate) instantons, to certain
interactions in the local quantum mechanics of the D0-branes world-line. 
In this way we provide a number of non-trivial checks on the
D-brane construction of Yang--Mills instantons, as presented in 
\refs{\rwinst,\rdoug}.    
   
As Witten has first shown in ref.\ \rwittenM,
and further discussed in refs.\ \refs{\rlands,\rbrand,\rmberk,\rkapl}, 
 these configurations of NS, D4 and D6-branes
can be lifted to M-theory where one obtains the exact solution of the theory
through the Seiberg-Witten curves which are nothing else than the 
algebraic curves defining the geometry of the branes in the internal,
i.e.\ $(x^4,x^5,x^6,x^7,x^8,x^9,x^{10})$, space.   Of course, the 
solution, being exact, contains all the contributions from the field theory
instantons even if in the Type-IIA model only NS, D4 and D6-branes appear. 
Indeed, in lifting the Type-IIA configuration to M-theory, the NS-branes 
become the M-theory fivebrane, the D4-branes are also M-theory fibranes 
this time with one world-volume direction along the $x^{10}$ axis. The
D6-branes in M-theory are ``Kaluza-Klein monopoles'' or a 
multi-Taub-NUT space and they appear as a background internal space on 
which the M-theory fivebranes propagate. Thus the NS-branes and the D4-branes
become in M-theory a single fivebrane whose algebraic equation is the
Seiberg-Witten curve of the model. From M-theory point of view, the 
intersection between NS-branes and D4-branes is not ``singular'' anymore, 
and its resolution is described by the Seiberg-Witten curve. 

Although our discussion applies strictly
 to the regime of weak type-IIA coupling,
it is interesting to ask 
what happens of the D0-branes we have been discussing in this paper, in
the lifting to M-theory.  A free D0-brane of the type-IIA theory  
becomes a graviton Kaluza-Klein                 
 state carrying momentum along the circle in the eleventh direction,  
and  electrically charged with respect to the $U(1)$ gauge field 
corresponding to the rotations around this circle. As discussed in
 refs.\ \refs{\rwitqcd,\rmqcd}, the physics of such 
free D0-branes seems to have nothing to do with the field theory
SYM models we are interested in. Whereas in the M-theory picture it could be
not so simple to decouple these modes, in the Type-IIA construction we can
just declare that we do not consider free D0-branes. In this way, we are
making an explicit distinction between the free D0-branes, and the
D0-D4 branes bound states. As stated in section 2, this distinction is 
natural from the point of view of the type-IIA configuration, since 
there is no non-contractible circle where we could wrap a free D0-brane
world-line to 
yield a semiclassical instanton contribution. On the other hand, the
world-line of the bound D0-brane is just the euclidean world-line
of the D4-brane with which it is solidary. In other words, once the
D0-branes are bound to the D4-branes, they must be considered as part
of the dynamical data on the D4-branes world-volume.

Thus, the  only D0-branes
of interest are those which are bound to at least two D4-branes and are not
allowed to become free. In the internal space, the D0-branes just completely
adhere to the D4-branes since they differ only in the space-time directions
where the D0-branes are at a point. In lifting to M-theory, only the 
internal space gets modified and from this point of view the bound 
D0-branes just completely merge in the D4-branes and they become part
of the resulting fivebrane. Actually one could see this from the 
opposite point of view: in extending the D4-branes to become an M-theory
fivebrane wrapped around the eleventh circle, 
one is effectively dressing them up with bound D0-branes.

\newsec{Acknowledgements}
This work is partially supported by the European Commission TMR programme
ERBFMRX-CT96-0045 in which A.P.\ is associated to the Milano University.

\listrefs

\vfill\eject
\bye